\documentclass[12pt,a4paper]{article}

\usepackage{comment}
\usepackage{tikz}

\usepackage{caption}

\usepackage{mathtools}

\usepackage[mathscr]{euscript}

\usepackage{slashed}

\usepackage{amsmath}

\date{}

\title{\textbf{
Conic Constrained Particle Quantization
within the DB, FJBW and BRST Approaches
}}
\author{ \textbf{
Gabriel D. Barbosa$^{a}$ and
Ronaldo Thibes$^{b}$}
\\\\
\textit{$^{a}$\small{Escola de Qu\'\i mica, Universidade Federal do Rio de Janeiro}}\\
\textit{\small{Rio de Janeiro, C.P. 68542, Brazil}}\\\\
\textit{$^{b}$\small{Universidade Estadual do Sudoeste da Bahia}}\\
\textit{\small{Rodovia BR 415, Km 03, S/N -- Itapetinga, Bahia, Brazil}}
 }

\usepackage{graphicx}
\usepackage{amssymb,amsfonts,amssymb,amsthm}
\usepackage{amsmath}

\begin{document}

\maketitle

\abstract{We consider a second degree algebraic curve describing a general conic constraint imposed on the motion of a massive spinless particle.
The problem is trivial at classical level but becomes involved and interesting in its quantum counterpart with subtleties in its symplectic structure
and symmetries.  The model is used here
to investigate quantization issues related to the Hamiltonian constraint structure, Dirac brackets, gauge symmetry and BRST transformations.
We pursue the complete constraint analysis in phase space and perform the Faddeev-Jackiw
symplectic quantization following the Barcelos-Wotzasek iteration program to unravel the fine tuned and more relevant aspects of the constraint
structure. A comparison with the longer usual Dirac-Bergmann algorithm, still more well established in the literature, is also presented.
While in the standard DB approach there are four second class constraints, in the FJBW they reduce to two.  By using the symplectic potential
obtained in the last step of the FJBW iteration process we construct a gauge invariant model exhibiting explicitly its BRST symmetry.
Our results reproduce and neatly generalize the known BRST symmetry of the rigid rotor showing that it constitutes a particular case of a broader class of theories.
}

\section{Introduction}\label{int}
Is the rigid rotor a {\it gauge theory} or not?  By a rigid rotor we simply mean a particle moving on a circle with constant speed.
In this paper we concur with the fact that such a question does not make sense at all because, of course, the same physical model
may admit more than one mathematical description (for instance, different Lagrangians) and both answers could be acceptable.
By the same token the usual electromagnetism, regarded as one of the best prototypes of a very successful {\it gauge theory},
may or may not enjoy gauge freedom depending on how one starts its defining description.  These are old long known
facts which sometimes are not stressed enough nor recalled during the physicist daily research labor battles. 
We mention and concede, on the other hand, that the quest for gauge theories with their undeniable beauty appeal has
driven generations of physicists since last century for many decades
and is likely to continue for 
a good amount of next ones.
We do not take sides here but rather try to
maximize the profit from the many different views one can get by describing the same physical system with different mathematical models.
By starting with ideas from a constrained particle moving on a circle we hope to finish this paper by adding one more $0+1$ interesting
gauge model to the theoretical physicist toolbox.

In 1987, Nemeschansky, Preitschopf and Weinstein published {\it A BRST Primer} \cite{Nemeschansky:1987xb} where the simple
model of a particle constrained to move on a circle was quantized with Becchi-Rouet-Stora-Tyutin (BRST) symmetry
\cite{Becchi:1974xu, Becchi:1974md, Tyutin:1975qk}.  At first the model was introduced and discussed
for pedagogical purposes, comparing its gauge and BRST issues with those of more robust field theory ones such as QED and QCD.  It is
in fact interesting to see field (and string) theory advanced concepts appearing in a natural way in such a simple quantum mechanical
model.
The main focus of \cite{Nemeschansky:1987xb} was on BRST symmetry itself within the scope of field theory -- analogies between the Lagrange
multiplier for the circle path constraint and the $A_0$ component of the electromagnetic or Yang-Mills gauge field and comparisons with the correct gauge-fixing processes were deeply explored.
Naturally, in order to permit such analogies the particle moving on a circle problem had somehow to be described in a gauge invariant way -- this was done with the aid of a first-order Lagrangian.
In that paper, however, neither the role of Dirac brackets nor symplectic quantization methods were discussed.
To circumvent the use of Dirac brackets, the authors of \cite{Nemeschansky:1987xb}, taking advantage of the circular symmetry, rely on the use of polar coordinates and, to obtain gauge
invariance, discard the term proportional to the radial momentum $p_r^2$ from the Hamiltonian.

Some years later, in the beginning of the current century, Scardicchio \cite{Scardicchio:2002} first considered the problem of a particle constrained on a circle from the Dirac-Bergmann (DB)
\cite{Dirac:1950pj, Anderson:1951ta}
point of view performing a careful constraint analysis.  Actually, in \cite{Scardicchio:2002}, the author compares two different approaches, in the first one eliminating completely the constraint
by using polar coordinates and promoting the rotating angle $\theta$ and its conjugated momentum $p_\theta$ to Hermitian operators satisfying ordinary canonical commutation relations -- this
turns out to be 
possible due to the natural match between the circular constraint and polar coordinates which renders the radial one trivially constant.
In his second approach, in the same paper, Scardicchio calculates the Dirac brackets associated to the cartesian coordinates maintaining a whole set of
four second class constraints at quantum level.
These four constraints come from the standard DB algorithm of imposing time conserving consistency conditions. 
Since the main focus in \cite{Scardicchio:2002} is the quantization of the circle constrained particle itself, the author does not discuss gauge nor BRST issues but
rather proceeds to the Hilbert space to construct quantum operators satisfying the Dirac bracket algebra.

More recently, Nawafleh and Hijjawi \cite{Nawafleh:2013} generalized Scardicchio's contribution to a particle constrained to move on an elliptical path.
The four second-class constraint structure remains and the corresponding Dirac brackets for the elliptical path are 
straightforwardly generalized in terms of the two ellipse defining axis.
There were some minor technical inaccuracies in the results for the Dirac brackets presented in \cite{Nawafleh:2013}.
Three years ago another interesting pedagogical paper appeared
-- {\it Dirac Bracket for Pedestrians} -- by M. K. Fung \cite{Fung:2014}.  Fung discusses the essential ideas behind Dirac
brackets for singular systems using as main example the particle constrained on a circle.  Actually Fung does not follow blindly DB's algorithm but with
a clever method shows how to obtain the Dirac brackets in that case by simply inverting a two-by-two matrix.  He works with a reduced set of only two
constraints resembling in a manner the Faddeev-Jackiw approach which will be discussed here for a general conic.  In \cite{Fung:2014}, Fung also handles the elliptical case
exhibiting the corresponding correct Dirac brackets.
It can be seen though that, in the ellipse generalization worked out in \cite{Nawafleh:2013, Fung:2014}, it is not as direct a matter to eliminate
the constraint by using polar coordinates nor obtain a gauge symmetry as in the circle case.  Towards that direction we understand that it is more natural to proceed
with cartesian coordinates as we shall show explicitly exhibiting a gauge invariance for a generic conic path.

Still regarding the BRST symmetry originally proposed by Nemeschansky, Preitschopf and Weinstein \cite{Nemeschansky:1987xb}, some recent works have appeared
in the literature concerning a toy model for Hodge theory \cite{Gupta:2009dy}
and exploring the supervariable formalism \cite{Shukla:2014spa}.
However, up to now, the relying on polar coordinates and circular symmetry seems to have been mandatory.
With that motivation in mind,
it is one of our main goals here to show how to proceed without circular symmetry nor the necessity of parametric coordinates matching the constraint, such
as the polar ones for the circle, and still obtain a generalized BRST symmetry.

In this context,
in the present article we propose a generalization of all the previous discussed ideas to an arbitrary conic described by a second degree algebraic curve.
In section {\bf 2} we introduce the model describing a particle constrained to move on the referred conic and go through its classical equations of motion.
Because of the simplicity of the constraint the model is readily shown to be integrated by using inverse elliptic functions -- 
we exemplify the general solution in the ellipse case.  In section {\bf 3} the canonical quantization
of the model
is worked out by using the standard DB approach.  Since the DB method is widely established in the physical literature we go through its calculational steps at a somewhat
rapid pace in order to
save space for the less known and more succinct symplectic Faddeev-Jackiw (FJ) one. In section {\bf 4} we perform the detailed calculations
concerning the FJ procedure \cite{Faddeev:1988qp} -- based on the analysis of the one-form associated to the kinematics term of the first-order Lagrangian.  We follow the iteration program proposed by Barcelos-Neto and Wotzasek \cite{BarcelosNeto:1991kw} and confirm that the FJ
brackets agree with Dirac's.   In section {\bf 5} we present a new gauge model which generalizes \cite{Nemeschansky:1987xb}
for the arbitrary conic described in cartesian coordinates.  After performing the gauge-fixing, by introducing the usual ghost Grassmannian coordinates, a remaining BRST symmetry is explicitly shown to survive.
All previously discussed particular cases published so far in the literature are then recovered by choosing specific values for the coefficients of the arbitrary conic.

\section{Classical Lagrangian Analysis}
In this section we introduce the simple model to be considered throughout the paper defined by the Lagrangian
\begin{equation}\label{L}
 L(x,y,z,\dot{x},\dot{y},\dot{z}) = \frac{1}{2} m (\dot{x}^2+\dot{y}^2) + zT(x,y)
\,.
\end{equation}
At first $T(x,y)$ can describe an arbitrary algebraic curve but for practical purposes
and the sake of comparison with the current literature, in this article, we shall focus on a generic
quadratic function given by
\begin{equation}
 T(x,y) = \frac{1}{2} Ax^2 + \frac{1}{2} B y^2 + Cxy +Dx +Ey +F
\end{equation}
with the constants $A,\dots,F$ denoting real parameters characterizing a specific conic.
From the classical point of view the model describes a particle constrained to move on the two-dimensional plane curve $T(x,y)$.
  The third variable $z$ plays the role of a Lagrange
multiplier\footnote{We stress however that all three dynamical variables $(x,y,z)$ are treated in the present formalism at exactly the same level.} naturally enforcing the constraint $T(x,y)=0$.  Although extremely simple, the model (\ref{L}) describes a singular system
from the Dirac-Bergmann (DB) point of view \cite{Dirac:1950pj, Anderson:1951ta} and exhibits interesting features in its quantum
version to be discussed in the forthcoming sections.

By demanding stationarity of the corresponding action with respect to arbitrary variations in the coordinates $x,y$ and $z$, fixed as usual
at the boundary of the time interval, the associated 
Euler-Lagrange (EL) equations of motion read
\begin{eqnarray}\label{EL}
 m\ddot{x} - z T_x &=& 0\,, \nonumber\\
 m\ddot{y} - z T_y &=& 0\,, \nonumber\\
 T(x,y)&=& 0
\,,
\end{eqnarray}
where we have introduced the handy notation
\begin{equation}\label{Tx}
 T_x\equiv\frac{\partial T}{\partial x}=Ax+Cy+D
\,,
\end{equation}
and
\begin{equation}\label{Ty}
 T_y\equiv\frac{\partial T}{\partial y}=By+Cx+E
\,.
\end{equation}
Relations (\ref{EL}) comprise a
system of coupled ordinary differential equations
for the three unknown functions $x(t)$, $y(t)$ and $z(t)$ --
actually a very easy one to solve because $z$ has no dynamics and the last one establishes a direct functional
relation between $x$ and $y$ without any derivatives.
After eliminating $z$, the first two EL equations lead to
\begin{equation}\label{xyEL}
 \ddot{x} T_y = \ddot{y} T_x
\,.
\end{equation}
The elimination of one further dynamical variable, let us say $y(t)$, can be done using the last EL equation (\ref{EL}).
Namely, we can
solve for $y$ as a function of $x$
\begin{equation}\label{fx}
 0=T(x,y)\rightarrow y=f(x)
\end{equation}
and obtain $\dot{y}(x,\dot{x})$ and $\ddot{y}(x,\dot{x},\ddot{x})$ by direct time derivation as well.
Since $T(x,y)$ describes a conic, actually $y$ may be obtained as a doubly degenerated function of $x$, corresponding to two ramifications.
Then the analysis can be split into the two possibilities in the specific case.
Back to (\ref{xyEL}) with (\ref{fx}), we
obtain a second order ordinary differential equation
\begin{equation}\label{Mx}
 P(x,\dot{x},\ddot{x}) = 0
\,.
\end{equation}
In a similar manner for $y$ we have
\begin{equation}\label{Ny}
 Q(y, \dot{y}, \ddot{y}) = 0
\,.
\end{equation}
Finally by using the constant of motion $v^2=\dot{x}^2+\dot{y}^2$ the order of (\ref{Mx}) and (\ref{Ny}) can be reduced and directly integrated.
In this way we achieve the general solution of the Euler-Lagrange equations (\ref{EL}).

As an illustrative example, let us consider an ellipse centered at the origin described by
\begin{equation}
 T(x,y) = \frac{x^2}{a^2}+\frac{y^2}{b^2} - 1
\,.
\end{equation}
Solving for $y$ as a function of $x$, performing time derivatives, substituting into (\ref{xyEL}) and performing algebraic manipulations leads
to the expression
\begin{equation}
 P(x,\dot{x},\ddot{x}) = \left[
a^6 + (a^2b^2-2a^4)x^2 - (b^2-a^2)x^4 \right]
\ddot{x} + a^2b^2x\dot{x}
\end{equation}
for the function defined in equation (\ref{Mx}) resulting in a second order ordinary differential equation for $x(t)$.  Multiplication by the integration factor
\begin{equation}
 I = \frac{\dot{x}}{{(a^2-x^2)}^2}
\end{equation}
permits then to rewrite (\ref{Mx}) in the present case as
\begin{equation}
 \frac{a^4+(b^2-a^2)x^2}{a^2-x^2}{\dot{x}}^2 = a^2 v^2
\end{equation}
where $v$ is a first integration constant.  Then a second integration leads the solution
\begin{equation}
 \int_{x_0}^{x(t)} du \left[
\frac{a^4+(b^2-a^2)u^2}{a^2(a^2-u^2)}
\right]^{1/2} = vt
\,.
\end{equation}
A similar expression can be obtained for $y(t)$ performing the same previous steps exchanging the roles of $x$ and $y$.

\section{Dirac-Bergmann 
Approach}
Having performed the classical Lagrangian analysis of the conic constrained particle in the last section we wish now to pursue its quantization.
The most direct method is the canonical quantization relying on a Hamiltonian basis.  It happens however that, since we are dealing with a constrained
system, some care must be taken.
In this section we discuss the application to (\ref{L}) of the well-known Dirac-Bergmann (DB) algorithm \cite{Dirac:1950pj, Anderson:1951ta} -- proper
and suited for constrained systems.
Since this is a common standard procedure we present only the main results which can be easily checked by the reader.
Nice traditional reviews of the DB formalism can be found for instance in \cite{Sundermeyer:1982gv, Henneaux:1992ig}.
Starting from (\ref{L}), after introducing the canonical momenta $(p_x,p_y,p_z)$, the canonical Hamiltonian can be straightforwardly calculated as
\begin{equation}
 H = \frac{p_x^2+p_y^2}{2 m}-zT(x,y)
\end{equation}
being well defined only in the primary constraint surface
\begin{equation}\label{phi1}
 \phi_1 \equiv p_z
\,.
\end{equation}
Then time conservation of the primary constraint $\phi_1$ and subsequent ones leads to
the following chain of three more secondary constraints
\begin{eqnarray}
   \phi_2 &\equiv& T(x,y) \,, \nonumber \\
   \phi_3 &\equiv& \displaystyle\frac{p_x}{m} T_x + \displaystyle\frac{p_y}{m} T_y \,, \nonumber \\
   \phi_4 &\equiv&  \frac{p_x^2}{m^2} A+ \frac{p_y^2}{m^2} B + 2 C\frac{p_x p_y}{m^2}+ \frac{z}{m} (T_x^2+T_y^2)
\,.
\end{eqnarray}
We recall $T_x$ and $T_y$ have been defined in equations (\ref{Tx}) and (\ref{Ty}).
Further time conservation of the last constraint $\phi_4$ does not lead to new ones but rather determines the Lagrange multiplier associated
to the primary constraint $\phi_1$ in the so-called primary Hamiltonian \cite{Sundermeyer:1982gv, Henneaux:1992ig}.
We denote the DB constraints collectively as $\phi_a$ with $a=1,\dots,4$.
By calculating the usual Poisson brackets among all four $\phi_a$
we form the {\it constraint matrix} written in closed form as
\begin{equation}\label{CM}
C_{ab}\equiv
[
 \phi_a,\phi_b 
]
=\left(
\begin{array}{rc}
{\bf 0}_2 & {\bf B} \\
{\bf -B} & {\bf C} \\
\end{array}
\right)
\end{equation}
where 
\begin{equation}{\bf B}=
\frac{1}{m}
\left(
\begin{array}{cc}
  0 & {-T_x^2-T_y^2}{} \\
  {T_x^2+T_y^2}{} & 2(U_x T_x+ U_yT_y) \\
 \end{array}
\right)
\end{equation}
and
\begin{equation}
{\bf C}=\left(
\begin{array}{cccc}
  0 & K \\
  -K & 0 \\
\end{array}
\right)
\end{equation}
with
\begin{eqnarray}
 K &\equiv& \frac{2}{m^2} \Big[ m (U_x^2+U_y^2)-z T_x (A T_x+C T_y)- z T_y (C T_x+B T_y) \Big] 
\,,
\end{eqnarray}
\begin{equation}\label{Ux}
 U_x\equiv \frac{1}{m} (Ap_x+Cp_y)
\,,
\end{equation}
and
\begin{equation}\label{Uy}
 U_y\equiv \frac{1}{m}(Cp_x+Bp_y)
\,.
\end{equation}
Still further notation, $
{\bf 0}_2$ in (\ref{CM}) denotes the constant null two-by-two matrix.
The definitions of $U_x$ and $U_y$ above will be also useful in the next section where they will appear more naturally\footnote{Here they can be understood
as coming from derivatives of $\phi_3$ with respect to $x$ and $y$.}.

The determinant of the four-by-four constraint matrix (\ref{CM}) can be readily calculated as
\begin{equation}\label{CMdet}
 \det C_{ab} = 
 \frac{{\left( T_x^2+T_y^2 \right)}^4}{m^4}
\end{equation}
and being non-null ascertain the second class nature of all 
four constraints $\phi_a$.  That means there is no gauge freedom for the model in
its original form (\ref{L}) and the Dirac brackets can be straightforwardly calculated after inverting (\ref{CM}).  The non-null brackets involving
the $x$ and $y$ variables and their conjugated momenta turn out to be
\begin{eqnarray}
 \left[x,p_x\right]^*&=&\frac{T_y^2}{T_x^2+T_y^2} \,,\nonumber \\
 \left[x,p_y\right]^*&=&-\frac{T_x T_y}{T_x^2+T_y^2}         = \left[ y,p_x\right]^*                \,,\nonumber \\
 \left[y,p_y\right]^*&=& \frac{T_x^2}{T_x^2+T_y^2} \,,\nonumber \\
 \left[p_x,p_y\right]^*&=& \frac{m(U_x T_y-U_y T_x)}{T_x^2+T_y^2} 
\label{DBs}
\,, 
\end{eqnarray}
while those related to the $z$ variable read
\begin{eqnarray}
 \left[x,z\right]^*&=&\frac{2 T_y}{m \left(T_x^2+T_y^2\right)
^2} \left\{p_x(CD-AE) 
 + p_y (DB-CE)+(AB-C^2)(xp_y-yp_x)
\right\} \,,\nonumber \\
 \left[y,z\right]^*&=&\frac{2 T_x }{m \left(T_x^2+T_y^2\right)
^2} \left\{ p_x \left(A E -D C\right) +p_y(CE-DB) - \left( A B - C^2  \right)(xp_y - yp_x) \right\} \,,\nonumber \\
\left[z,p_x\right]^*&=& \displaystyle\frac{2}{ \left(T_x^2+T_y^2\right)
^2}\left\{
mU_x (U_xT_x+ U_y T_y) 
-z\left( AT_x+CT_y \right) \left( T_x^2 + T_y^2 \right)
\right.\nonumber \\ 
  &+&  
\left.
T_x \left[z T_x (A T_x+C T_y)+ z T_y (C T_x+B T_y)- m\left(U_x^2-U_y^2\right)   \right] 
\right\} \,,\nonumber \\
\left[z,p_y\right]^*&=& \displaystyle\frac{2}{ \left(T_x^2+T_y^2\right)
^2} 
\left\{
mU_y (U_xT_x+U_y T_y) 
-z\left( CT_x+BT_y \right) \left( T_x^2 + T_y^2 \right)
\right.\nonumber \\
  &+& 
\left.
T_y \left[ z T_x (A T_x+C T_y)+ z T_y (C T_x+B T_y)- m\left( U_x^2- U_y^2\right)  \right] 
\right\}
\,.
\end{eqnarray}
We use a star to denote Dirac brackets in order to distinguish from the ordinary Poisson brackets.
The Dirac bracket of $p_z$ with any other phase space function vanishes identically -- as it should, because of (\ref{phi1}).

As an interesting situation, a special case considered recently in \cite{Nawafleh:2013, Fung:2014}, we mention an elliptical trajectory centered at the origin with major and minor axis respectively
$a=2/A$ and $b=2/B$.  In this case we have
\begin{equation}
 C=D=E=F=0
\,,
\end{equation}
the brackets (\ref{DBs}) reduce straightforwardly to
\begin{eqnarray}\label{ellipseDBs}
 \left[x,p_x\right]^*&=&\displaystyle\frac{a^4 y^2}{b^4 x^2+a^4 y^2} \,, \nonumber \\
 \left[x,p_y\right]^*&=&-\displaystyle\frac{a^2 b^2 x y}{b^4 x^2+a^4 y^2}  =\left[ y,p_x\right]^* \,,\nonumber \\
 \left[y,p_y\right]^*&=& \displaystyle\frac{b^4 x^2}{b^4 x^2+a^4 y^2} \,,\nonumber \\
 \left[p_x,p_y\right]^*&=& \frac{a^2 b^2 
(yp_x-xp_y)
}{b^4 x^2+a^4 y^2}\,,
\end{eqnarray}
and we have omitted the ones involving the non dynamical variable $z$.
We take the opportunity to point out missing factors of $a^2$ and $b^2$ in the denominators of the corresponding expressions for $[x,p_y]^*$ and $[y,p_x]^*$
in \cite{Nawafleh:2013}. In fact these two brackets must be dimensionless. Moreover, for a circle with radius $r\equiv b=a$ 
the brackets (\ref{ellipseDBs}) completely agree with those previously published by Scardicchio \cite{Scardicchio:2002} and Fung \cite{Fung:2014}.

In the next section we shall show how to achieve the same results concerning the general Dirac brackets (\ref{DBs}) with the modern symplectic constraint treatment
due to Faddeev and Jackiw.  The FJ approach will prove to be more economical and direct to the point, needing only half the current DB constraints.

\section{Symplectic Quantization}
In a ingenious paper, concentrated on first-order Lagrangians, Faddeev and Jackiw \cite{Faddeev:1988qp} have inaugurated a subtle and simpler form of
treating singular constrained systems.  In \cite{Faddeev:1988qp} Dirac's original constraint classification is criticized and it is shown that
in certain cases the DB algorithm produces unnecessary artificial constraints.  The use of symplectic methods is used to unravel the
there denominated {\it true constraints}, bypassing the trivial DB ones.  Once established the initial {\it true constraints}, Faddeev and Jackiw
consider either their elimination by coordinate transformations or the application of the DB algorithm to an intermediate step modified Lagrangian.
Building on the original FJ work, Barcelos-Neto and Wotzasek \cite{BarcelosNeto:1991kw} proposed an iteration algorithm which inserts consistency constraint
time conservation into FJ's symplectic approach dispensing completely further need of the DB algorithm thus making the method self-consistent -- this
is what we call here the FJBW iteration procedure.
In this section we apply the FJBW symplectic algorithm to the model (\ref{L}) reducing the number of constrained from four to two and obtain
the FJ brackets direct from the last step iterated Lagrangian one-form.

As usual, in the symplectic formalism \cite{Faddeev:1988qp, BarcelosNeto:1991kw}, with the aid of additional auxiliary variables, one reduces the Lagrangian to first-order
and writes
\begin{equation}\label{L0}
 L^{(0)} = a_i^{(0)}(\xi^{(0)})\dot{\xi}_i^{(0)} - W^{(0)}(\xi^{(0)})
\end{equation}
where $\xi^{(0)} = ( \xi^{(0)}_i )$ represents the set of symplectic variables with the index $i$ running
through all of them.  The upperscript ${}^{(0)}$ is used because of the natural iterative
procedure of the symplectic method as later on new iterated Lagrangians $L^{(k)}$, with $k>0$, are to be calculated.
Further $a_i^{(0)}$ and $W^{(0)}$  in (\ref{L0}) represent respectively the canonical one-form and the
initial zero-order
symplectic potential collecting all terms in the first-order Lagrangian without time derivatives.
In the present case, by introducing two auxiliary variables $p_x$ and $p_y$ we may write (\ref{L}) in first-order as
\begin{equation}\label{L0conic}
 L^{(0)} = p_x\dot{x} + p_y\dot{y} - \frac{1}{2m}(p_x^2+p_y^2)+zT(x,y)
\end{equation}
and we have the initial five symplectic variables $\xi^{(0)} = (x,y,z,p_x,p_y)$.
By comparison with (\ref{L0}), 
the starting symplectic potential is given by the expression
\begin{equation}\label{W0}
 W^{(0)}(x,y,z,p_x,p_y) = \frac{1}{2m}(p_x^2+p_y^2) - zT(x,y)
\,.
\end{equation}

The EL equations of motion can be written in the current symplectic formalism as
\begin{equation}\label{ELsymp}
 f_{ij}^{(0)} \dot{\xi}^{(0)}_j = \frac{\partial W^{(0)}}{\partial \xi_i^{(0)}}
\,,
\end{equation}
with the symplectic two-form
\begin{equation}
f_{ij}^{(0)}\equiv \frac{\partial a^{(0)}_j}{\partial \xi_i^{(0)}} -  \frac{\partial a_i^{(0)}}{\partial \xi_j^{(0)}} 
\,.
\end{equation}
Note that here the EL equations are first order differential equations, a natural bonus which comes as a result of introducing more variables.
In order to solve (\ref{ELsymp})
for the velocities $\dot{\xi}_{ij}^{(0)}$,
it is necessary to check for the reversibility of $f_{ij}^{(0)}$.  In the current model, described by Lagrangian (\ref{L0conic}), $f_{ij}^{(0)}$ is clearly not invertible
because the variable $z$ does not show up in the kinetic part of (\ref{L0}) and therefore has no dynamics.
Naturally this is a fingerprint of constrained systems being equivalent, in Dirac's standard procedure, to a non-null
Hessian for the corresponding second order version (\ref{L}).
In general, the singularity of $f_{ij}^{(0)}$ implies the existence of zero modes $v_\alpha^{(0)}$ leading to the kinematic
symplectic constraints
\begin{equation}\label{Omega0}
 \Omega_\alpha^{(0)} \equiv (v_\alpha^{(0)})_i \frac{\partial W^{(0)}}{\partial \xi_i^{(0)}}
\,.
\end{equation}
In the model (\ref{L}) we obtain only one zero mode for $f^{(0)}$, namely
\begin{equation}\label{v0}
 v^{(0)} = (\,0\,,\,\,\,0\,,\,\,\,-1\,,\,\,\,0\,,\,\,\,0\,)
\,,
\end{equation}
corresponding to the obvious kinematic constraint
\begin{equation}\label{T}
 \Omega^{(0)} = T(x,y)
\end{equation}
obtained directly from (\ref{Omega0}) applied to (\ref{W0}).
The arbitrary negative sign in (\ref{v0}) stands only for convenience.
The general idea of the FJBW procedure amounts to introducing Lagrange multipliers to transfer the constraints (\ref{Omega0}) from the kinematic
to the symplectic potential sector by redefining all involved quantities through a iteration process until one renders the symplectic
matrix $f^{(k)}_{ij}$ invertible at some finite step $k$.  Therefore, for the first FJBW iteration step in our model, we introduce
a Lagrange multiplier $\lambda$ and impose time conservation of (\ref{T}) leading to a first iterated
Lagrangian
\begin{equation}\label{L1}
 L^{(1)} = p_x \dot{x} + p_y\dot{y} + T(x,y)\dot{\lambda} - W^{(1)}(p_x,p_y)
\end{equation}
with
\begin{equation}\label{W1}
 W^{(1)}(p_x,p_y) = W^{(0)}{\Big\vert}_{T(x,y)=0} = \frac{1}{2m}(p_x^2+p_y^2)
\,.
\end{equation}
Since now neither $z$ nor its time derivative appear in (\ref{L1}) anymore we may drop it from the first iterated set of symplectic variables
and define
\begin{equation}
 \xi^{(1)}\equiv(x,y,p_x,p_y,\lambda)
\,.
\end{equation}
For notational purposes we introduce the canonical symplectic matrix
\begin{equation}
 \mathbf{J}_{4} \equiv \left( \begin{array}{cr} {\bf 0}_2&-\mathbf{I}_{2}\\\mathbf{I}_{2}&{\bf 0}_2 \end{array} \right)
\end{equation}
where $\mathbf{I}_{2}$ denotes the identity two by two matrix.
From the kinetic part of (\ref{L1}), using $\mathbf{J}_4$,
we may write the first-iterated canonical two-form as
\begin{equation}\label{f1}
 f^{(1)} = \left(
\begin{array}{cc}
 \mathbf{J}_{4}&\begin{array}{c}T_x\\T_y\\0\\0\end{array}\\
 \begin{array}{cccc}-T_x&-T_y&0&0\end{array}&0
\end{array}
\right)
\end{equation}
which, being antisymmetric and odd-dimensional, is necessarily singular thus requiring one more step in the symplectic FJBW algorithm.
Proceeding further this next step, note that (\ref{f1}) enjoys a zero-mode given by
\begin{equation}
 v^{(1)} = (\,0\,,\,\,\,0\,,\,\,\,T_x\,,\,\,\,T_y\,,\,\,\,1\,)
\end{equation}
which, similarly to (\ref{Omega0}), insures the constraint
\begin{equation}\label{U}
 U(x,y,p_x,p_y)\equiv
 (v^{(1)})_i\left(\frac{\partial W^{(1)}}{\partial\xi_i^{(1)}}\right)=
\frac{1}{m}\left( p_xT_x+p_yT_y \right)
\end{equation}
requiring a second Lagrange multiplier $\eta$ to form the second-iterated first-order Lagrangian
\begin{equation}\label{L2}
 L^{(2)} = p_x\dot{x} + p_y\dot{y} +T\dot{\lambda}+U\dot{\eta} - W^{(2)}
\,.
\end{equation}
The symplectic potential $W^{(2)}$ is obtained from (\ref{W1}) by imposing the constraint (\ref{U}) and can be written as
\begin{equation}\label{W2}
  W^{(2)} = \frac{1}{2m}\frac{{\left( T_xp_y-T_yp_x \right)}^2}{T_x^2+T_y^2}
\,.
\end{equation}
We remark that the choice of notation for the constraint (\ref{U}) is consistent with the previous equations (\ref{Ux}) and (\ref{Uy}) which are now justified
as
\begin{equation}
 U_x\equiv \frac{\partial U}{\partial x}
\end{equation}
and
\begin{equation}
 U_y\equiv \frac{\partial U}{\partial y}
\,.
\end{equation}

The symplectic two-form associated to (\ref{L2}), within the second iteration step set of variables
\begin{equation}\label{symvars2}
\xi^{(2)}\equiv(x,y,p_x,p_y,\lambda,\eta)
\,, 
\end{equation}
is a six-by-six matrix given by
\begin{equation}\label{f2}
 f^{(2)} =
\left(
\begin{array}{cc}
 \mathbf{J}_{4}&\begin{array}{cc}T_x&U_x\\T_y&U_y\\0&T_x/m\\0&T_y/m\end{array}\\
   \begin{array}{cccc}
      -T_x&-T_y&0&0\\
      -U_x&-U_y&-T_x/m&-T_y/m
   \end{array}
   &
   \begin{array}{cc}
      ~~0~~~&~0~~~\\
      ~~0~~~&~0~~~
   \end{array}
\end{array}
\right)
\,,
\end{equation} 
with determinant
\begin{equation}
\det (f^{(2)}) = \frac{{\left(T_x^2+T_y^2\right)}^2}{m^2}
 \,\,.
\end{equation}
Note the close similarly with (\ref{CM}) and its determinant (\ref{CMdet}) obtained by the standard DB algorithm.
The non-singularity of (\ref{f2}) shows that we have achieved the final step of the FJBW iteration procedure and the FJ brackets can be read directly from
its inverse.  Indeed, the inverse of (\ref{f2}) can be cast into the form
\begin{equation}\label{f2inv}
 {\left[ \frac{f^{(2)}}{T_x^2+T_y^2} \right]}^{-1} =
\begin{pmatrix}
 0&0&{\it T_y}^2&-{\it T_x}{\it T_y}&-{\it T_x}&0\cr 0&0&-
 {\it T_x}{\it T_y}&{\it T_x}^2&-{\it T_y}&0\cr -{\it T_y}^2&{\it T_x}
 {\it T_y}&0&-mT_{[x}U_{y]}&m{\it U_x}&-
 m{\it T_x}\cr {\it T_x}{\it T_y}&-{\it T_x}^2&mT_{[x}U_{y]}&0&m{\it U_y}&-m{\it T_y}\cr {\it T_x}&
 {\it T_y}&-m{\it U_x}&-m{\it U_y}&0&-m\cr 0&0&m{\it T_x}&m
 {\it T_y}&m&0\cr
\end{pmatrix}
\end{equation}
where
\begin{equation}
 T_{[x}U_{y]}\equiv T_x U_y - T_y U_x
\,.
\end{equation}
Considering the conventional symplectic variables order defined in (\ref{symvars2}), from the four first rows and columns entries of (\ref{f2inv}) we obtain the
following non-null
FJ brackets
\begin{equation}\label{xpx}
 [x,p_x]_{{}_{FJ}} = \frac{T_y^2}{T_x^2+T_y^2}\,,\,\,\,\,\,[y,p_y]_{{}_{FJ}} = \frac{ T_x^2}{T_x^2+T_y^2}\,,
\end{equation}
\begin{equation}\label{xpy}
 [x,p_y]_{{}_{FJ}}=-[y,p_x]_{{}_{FJ}}=-\frac{T_xT_y}{T_x^2+T_y^2}
\end{equation}
and
\begin{equation}\label{pxpy}
 [p_x,p_y]_{{}_{FJ}}=-m\frac{T_{[x}U_{y]}}{T_x^2+T_y^2}
\,.
\end{equation}
As previously claimed we see that the FJ brackets above perfectly match (\ref{DBs}) agreeing with the results obtained
by the standard DB procedure.

Once the algebra of (\ref{xpx}-\ref{pxpy}) among the dynamical variables $(x,y,p_x,p_y)$ has been obtained, the canonical quantization process goes
as usual by promoting them to quantum operators acting on an appropriate Hilbert space.  Operator ordering issues can be tackled by
imposing Hermicity as was done for instance in \cite{Scardicchio:2002} for the circle case.
Alternatively, the quantization may also be performed by functional methods.  Along this line,
we proceed in the next section to our main goal
of obtaining gauge and BRST invariance for our model.

\section{Gauge and BRST Symmetries}
After considering the canonical quantization of the conic constrained particle ({\ref{L}) described either by the DB's or FJBW's approaches, in this section,
we discuss the very same model from a gauge invariance principle point of view.
As usual for gauge systems we shall also exhibit a BRST symmetry \cite{ Becchi:1974xu, Becchi:1974md, Tyutin:1975qk} which survives even after the breaking of the gauge one resulting from a specific gauge fixing choice.
The quantization then can be achieved by functional integration techniques.
We recall that in the particular case of the rigid rotor around the origin, using polar coordinates, a similar analysis has been performed in \cite{Nemeschansky:1987xb}
whose main ideas we now generalize.

Instead of using (\ref{L}) we describe the same system by the first-order Lagrangian
\begin{equation}\label{Linv}
 L_{inv}=
 p_x\dot{x}+p_y\dot{y}
 -\frac{1}{2m}\frac{{\left( T_xp_y-T_yp_x \right)}^2}{T_x^2+T_y^2}
 +zT(x,y)
\end{equation}
where the third term
\begin{equation}
 -\frac{1}{2m}\frac{{\left( T_xp_y-T_yp_x \right)}^2}{T_x^2+T_y^2} \equiv -W^{(2)}
\end{equation}
comes from the second iterated FJBW potential (\ref{W2}).
The fine point here is that, for a given arbitrary time dependent function $\phi(t)$, the Lagrangian $L_{inv}$ enjoys the following gauge symmetry
\begin{equation}
 z\rightarrow z + \dot{\phi} \,,\nonumber
\end{equation}
\begin{equation}\label{ginv}
 p_x\rightarrow p_x+\phi T_x\,,\,\,\,\,\,p_y\rightarrow p_y + \phi T_y \,,
\end{equation}
as can be checked by inspection.  In fact, under (\ref{ginv}), $L_{inv}$ changes by the total derivative
\begin{equation}
 \Delta  L_{inv}= \frac{d}{dt} \left( \phi T \right)
\end{equation}
and the corresponding time-integrated action remains invariant.

At quantum level we may introduce two Grassmann variables $c$ and $\bar{c}$ and, for gauge-fixing purposes, an additional Nakanish-Lautrup variable $N$.
Then the original gauge symmetry gives rise to the following BRST transformations
\begin{equation}
 \delta x = \delta y = 0 \,,\,\,\,\,\, \delta z = \dot{c}\,,  \nonumber
\end{equation}
\begin{equation}
 \delta p_x = cT_x\,,\,\,\,\,\,\delta p_y = cT_y\,, \nonumber
\end{equation}
\begin{equation}\label{BRSTtfs}
 \delta c = 0\,,\,\,\,\,\,\delta \bar{c} = N\,,\,\,\,\,\, \delta N = 0 \,.
\end{equation}
As usual, we may also associate ghost numbers $1$ and $-1$ to $c$ and $\bar{c}$ respectively, as summarized in
the table below.
\captionsetup{labelformat=empty}
\begin{table}[ht]
\centering
{\small\begin{tabular}{lcccccccc} \hline\hline &$x$&$y$&$z$&$p_x$&$p_y$&$c$&$\bar{c}$&N\\
\hline
Grassmann parity&$0$&$0$&$0$&$0$&$0$&$1$&$1$&$0$\\
ghost number&$0$&$0$&$0$&$0$&$0$&$1$&$-1$&$0$
\\
\hline\hline
\end{tabular}}
\caption{Grassmann parity and ghost numbers} \label{table1}
\end{table}
The BRST operator $\delta$ holds odd Grassmannian parity, carries ghost number one and is nilpotent as can be checked from (\ref{BRSTtfs}).

Now for a satisfactory functional quantization process a gauge-fixing term must be added to
(\ref{Linv}) -- taking advantage of the nilpotency of the BRST operator we choose the BRST exact term
\begin{equation}\label{dLgf}
 L_{gf} = \delta\left[
\bar{c}\left(\dot{z} + M(x,y)(T_xp_x + T_yp_y ) + \frac{N}{2} \right)
\right]
\end{equation}
where $M(x,y)$ is a suitable function for a proper gauge-fix.  By applying (\ref{BRSTtfs}) we obtain explicitly
\begin{eqnarray}
 L_{gf} &=& N\left[ \dot{z} + M(x,y)(T_x p_x + T_y p_y ) \right] + \frac{N^2}{2} \nonumber\\
 &&- \bar{c}\left[ \ddot{c} + (T_x^2+T_y^2)M(x,y)c \right]\label{Lgf}
\,.
\end{eqnarray}

Once the gauge is fixed in a BRST invariant way, by
exponentiating the sum of (\ref{Linv}) and (\ref{Lgf}) we construct the quantum vacuum generating functional as
\begin{equation}\label{Z}
 Z = \int [d\mu] e^{-\frac{i}{\hbar}\int\left( L_{inv}+L_{gf} \right)dt }
\end{equation}
with functional integration measure
\begin{equation}
 [d\mu] = [dx][dy][dz][dp_x][dp_y][dc][d\bar{c}][dN]
\,.
\end{equation}
The total quantum action in the exponential argument of (\ref{Z}), given explicitly by
\begin{eqnarray}
 S 
 &=&\int dt \left\{ 
 p_x\dot{x}+p_y\dot{y}+N\dot{z}
 -\frac{1}{2m}\frac{{\left( T_xp_y-T_yp_x \right)}^2}{T_x^2+T_y^2}
 +zT(x,y)
 +\frac{N^2}{2} 
\right.\nonumber\\&&\left.~~~~~~~~~~
+ N\left[ \dot{z} + M(x,y)(T_x p_x + T_y p_y ) \right]
- \bar{c}\left[ \ddot{c} + (T_x^2+T_y^2)M(x,y)c \right]\
\right\}
\end{eqnarray}
is BRST invariant by construction and assures the functional quantization of the model.

With the usual coupling of external sources, (\ref{Z}) can be used to generate all Green's functions of the theory.
This turns out to be possible due to the fact that although we have BRST invariance we have fixed the gauge freedom with (\ref{Lgf}).
We have thus achieved our final
goal of describing the original system of a constrained conic particle at quantum level with explicitly BRST symmetry.

\section{Conclusion}
We have pursued the quantization of a particle constrained to live on a general conic path described by a second degree algebraic curve in cartesian coordinates.
We have gone through both canonical and functional techniques.  Concerning first the canonical quantization approach we have seen that the symplectic FJBW procedure
lead to the FJ brackets in a straightforward way needing only two constraints.  The FJ brackets obtained were shown to coincide with the usual Dirac ones obtained
from a more involved four constraints structure.  The functional quantization approach, on the other hand, has led us to consider a gauge invariant model
for the conic contained particle.  After the gauge-fixing, a BRST symmetry survived, mixing the original variables with the extra introduced ghost and
Nakanish-Lautrup variables.  The gauge model obtained largely generalizes the previous known ideas for the rigid rotor BRST symmetry.
We have shown that the rigid rotor BRST symmetry is not a peculiar coincidence relying on circular symmetry but is a particular case of the more
general model considered here and can also be realized with ordinary cartesian coordinates.

Concerning the ideas discussed in the Introduction, we have provided a simple and interesting $0+1$ quantum mechanical model which can be described
either without gauge symmetry, with a rich second class constraint structure generalizing the circle and ellipse cases, or in a gauge invariant way.  The gauge invariant version was constructed
from the last iterated FJBW symplectic potential and has lead to the usual BRST symmetry at quantum level, mixing bosonic and fermionic variables.

\end{document}